\documentclass[aps,twocolumn,showpacs,preprintnumbers,amsmath,amssymb,floatfix,prl,superscriptaddress]{revtex4}

\usepackage{graphicx}
\usepackage{epstopdf}
\usepackage{dcolumn}
\usepackage{bm}
\usepackage[usenames]{color}
\usepackage{units}
\usepackage{float}
\usepackage{comment}

\graphicspath{{./}{./figs/}}

\begin{document}


\title{Coupling ferroelectricity with spin-valley physics in oxide-based heterostructure}

\author{Kunihiko Yamauchi}
\affiliation{%
ISIR-SANKEN, Osaka University, 8-1 Mihogaoka, Ibaraki, Osaka, 567-0047, Japan }

\author{Paolo Barone}
\affiliation{%
Consiglio Nazionale delle Ricerche (CNR-SPIN), 67100 L'Aquila, Italy
}%

\author{Tatsuya Shishidou}
\affiliation{%
Department of Quantum Matter, ADSM, Hiroshima University, Higashihiroshima 739-8530, Japan 
}%

\author{Tamio Oguchi}
\affiliation{%
ISIR-SANKEN, Osaka University, 8-1 Mihogaoka, Ibaraki, Osaka, 567-0047, Japan }

\author{Silvia Picozzi}
\affiliation{%
Consiglio Nazionale delle Ricerche (CNR-SPIN), 67100 L'Aquila, Italy
}%

\date{\today}
\newcommand{\ba}{Ba$_{2}$CoGe$_{2}$O$_{7}$}
\begin{abstract}
The coupling of spin and valley physics is nowadays regarded as a promising route toward next-generation spintronic and valleytronic devices. In the aim of engineering functional properties for valleytronic applications, we focus on the ferroelectric heterostructure BiAlO$_{3}$/BiIrO$_{3}$, where the complex interplay among trigonal crystal field, layer degrees of freedom and spin-orbit coupling mediates a strong spin-valley coupling. Furthermore, we show that ferroelectricity provides a  non-volatile handle to manipulate and switch the emerging valley-contrasting spin polarization.
\end{abstract}


\maketitle


Technological advances in future electronics strongly rely on the investigation
of quantum degrees of freedom of electrons. Beside electron charge,
spin represents the most studied example, 
due to its obvious connection with magnetic information storage. Spin may be also actively manipulated in 
devices based on spin-polarized transport \cite{ziese_mag}, as 
in the prototypical spin-FET proposed by Datta and Das\cite{dattadas}. 
Relativistic spin-orbit coupling (SOC), providing a link between the electron spin and { orbital angular } momenta,  plays a central role in spin-polarization effects while preserving time-reversal symmetry. 
Such spin-polarization effects may appear even in centrosymmetric crystals\cite{zunger_2014,riley_natphys2014}, provided that the ions experiencing a strong SOC have an inversion asymmetric local environment, 
ultimately due to the localized character of SOC itself.  In the absence of bulk inversion symmetry,  SOC lifts the spin degeneracy, 
leading to spin-splitting phenomena known as Dresselhaus\cite{Dresselhaus1955} and Rashba\cite{BychovRashba1984} effects. In the latter case, typically realized at interfaces or surfaces where inversion symmetry is structurally broken, the spin polarization is induced by an effective $k$-dependent magnetic field, whose strength can be 
modulated 
by applying an external electric field\cite{dattadas,nitta1997}. Alternatively, it has been recently understood that bulk Rashba-like 
 effects can be realized in ferroelectric (FE) semiconductors
\cite{domenicoGeTe},  {additionally} providing a non-volatile functionality (ferroelectricity) which is tightly bound to spin polarization and, as such, allows for its full-electric control\cite{silviaFERSC}.

Other degrees of freedom have been recently addressed, most notably
valley pseudospin, labelling the degenerate energy extrema in {the} momentum
space, which could be used in valleytronic devices 
exploiting the valley index of carriers to process information\cite{beenakker_2007}.
The research in this field has been boosted by
the 
discovery of hexagonal 2D 
crystals\cite{novoselov_2005,chemrev2013}, such as graphene
and transition-metal dichalcogenide monolayers, displaying valleys at the corners
(K points) of the Brillouin zone. 
Various schemes have been proposed to generate valley currents 
in graphene,  using line defects\cite{Gunlyce_2011} or strain\cite{Jiang_2013}, and in biased dichalcogenide bilayers\cite{wu_nature2013}.
The detailed investigation of these materials has significantly contributed to the fundamental understanding of valley physics, leading to the identification of intrinsic 
physical properties associated with valley occupancy\cite{xu_rev2014, Xiao_2007, Xiao_2012}. 
Interestingly, the symmetry properties of valley pseudospin implies that valley-contrasting
physical quantities may emerge whenever inversion symmetry is broken 
\cite{xu_rev2014, Xiao_2007}.
In this respect,
 the possibility to permanently control the valley phenomenology via a non-volatile FE polarization would open the path to novel device paradigms, in analogy with the proposed FE Rashba semiconductors\cite{silviaFERSC}.

\begin{figure}[th]
\begin{center}
{\center{
\includegraphics[width=86mm, angle=0]{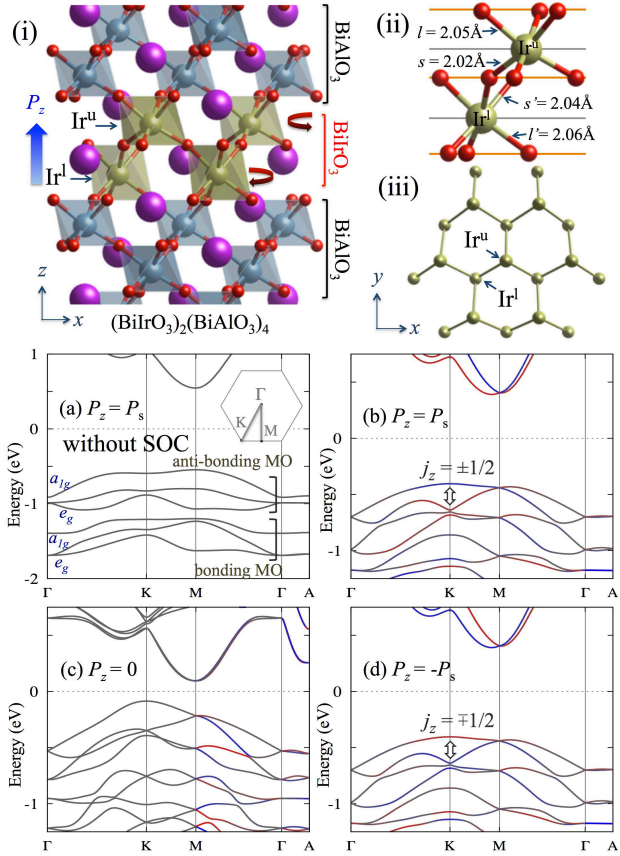}
\vspace {-0.6cm}
}}
\caption{\label{fig:crysband} 
(i) Multilayer structure of (BiIrO$_{3}$)$_{2}$(BiAlO$_{3}$)$_{4}$. 
Octahedral tilting and polarization direction are shown by arrows. (ii) Ir-O bond lengths in the 
bilayered structure and
(iii) top view of the corrugated honeycomb Ir lattice. 
Electronic bandstructure of FE (BiIrO$_{3}$)$_{2}$(BiAlO$_{3}$)$_{4}$ without SOC (a) and with SOC (b). 
With SOC, $P$ is tuned to $P=0$ in a PE parent structure (c) and to $P=-P_{\rm s}$ (d), where 
$P_{\rm s}$ is the calculated spontaneous  polarization. 
The $s_{z}$ polarization is highlighted by blue (up) and red (down) color.
Spin-splitting at VBM is indicated by block arrows.  
\vspace {-0.4cm}
}
\end{center}
\end{figure}

Most of the {well-known} FE materials are transition-metal oxides with perovskite structure. 
As recently pointed out, the hexagonal geometry can indeed be realized in perovskite heterostructures (ABO$_3$)$_n$/(AB'O$_3$)$_2$ grown along the [111] direction, where the B' ions in the (AB'O$_3$)$_2$ bilayer sit on a corrugated graphene-like honeycomb lattice\cite{nagaosa_2011,pardo_2013}, as shown in Fig. \ref{fig:crysband}. 
Recent technological developments\cite{norton_2004} allow now to prepare such artificial layered heterostructures
with atomic-scale precision\cite{rijinders_apl1997}, offering an unprecedented playground for the realization of oxide-based advanced electronics. 
In fact, these (111) superlattices could allow the integration of functional properties typically found in oxides\cite{nagaosatokura_nature2012} with the spin-valley physics, which is expected to emerge in the hexagonal layered structure. Specifically, the {\it ``layer degree of freedom''},  naturally emerging in the bilayer 
geometry, can be described as a {\it ``pseudospin''} up (down) labeling the state where the charge carrier is located in the upper (lower) layer, being therefore related to electrical polarization. Such a layer pseudospin is expected to couple with spin and valley pseudospin, possibly giving rise to exotic magnetoelectric effects
\cite{gong_2013}.
%

Since the spin-valley contrasting properties can emerge when inversion symmetry is broken, in this letter we focus on the quest for FE spin-valley physics in perovskite oxides.
As design criteria, a robust FE {system} 
with polarization perpendicular to the hexagonal bilayer is needed.
Since the BiFeO$_{3}$-type rhombohedral crystal structure, displaying a [111] polar axis, is considered as an ideal candidate to fulfill this criterion,
we choose the nonmagnetic isostructural BiAlO$_{3}$\cite{spaldin_BAO} 
 recently synthesized, with a measured polarization $P_s\approx27~\mu$C/cm$^2$ and a high critical temperature, $T_{c}> 520^{\circ}$\cite{son_BAO,zylberberg_bialo3}. 
Aiming at a sizable spin splitting in a nonmagnetic system, in this study we propose the low-spin Ir$^{3+}$ ($5d^6$) as a candidate B' ion, due to its large atomic SOC.
Density-Functional-Theory (DFT) calculations were performed using the VASP code with GGA-PBE potential\cite{vasp}. 

%
%
\begin{figure}[ht!]
\begin{center}
\vspace{-0.3cm}
{
\includegraphics[width=82mm, angle=0]{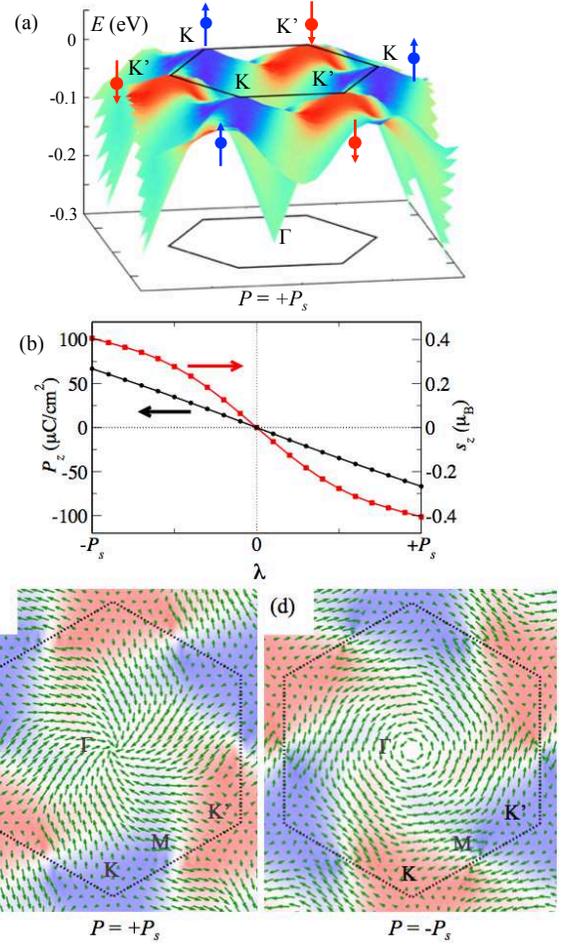}
\vspace {-0.4cm}
}
\caption{\label{fig:spinvalley} 
(a) Spin valleys of (BiIrO$_{3}$)$_{2}$(BiAlO$_{3}$)$_{4}$: $\pm s_{z}$ polarization projected on $E(k)$ curve of the VBM band. Top of $E_{\rm VBM}$ is set as 0 eV. 
(b) FE polarization $P_{z}$ and spin polarization $s_{z}$ of the VBM band at K point as a function of the polar structural distortion $\lambda$. 
(c), (d) Spin texture of the VBM band for +$P_s$ and -$P_s$ crystal structure, respectively: $s_{x}$ and s$_{y}$ components are shown by arrows, while $s_{z}$ component is plotted by blue ($+s_{z}$) and red ($-s_{z}$) color  of VBM band state. The 
first Brillouin zone is shown as a hexagon.  }
\vspace {-1.0cm}
\end{center}
\end{figure}

We first briefly summarize the outcome of our calculations for hypothetical bulk BiIrO$_{3}$. 
We found it to stabilize 
in the polar $R3c$ structure, 
with a calculated polarization $P_{s}$ = 70.6 $\mu$C/cm$^{2}$. 
As compared to the ideal perovskite 
structure, 
BiIrO$_{3}$ displays two main distortion modes, i.e., IrO$_{6}$ octahedral tilting and Bi-O polar distortion, the latter being responsible for the onset of ferroelectricity through the Bi lone-pair mechanism\cite{bifeo3.neaton}, alongside a compressive trigonal distortion. 
Ir$^{3+}$-5$d^{6}$ electrons stabilize in a low-spin state, whereas 
the trigonal distortion removes the degeneracy of fully-occupied $t_{2g}$ levels, which are split in a non-degenerate $a_{1g}$ and twofold degenerate $e'_g$ states. 

The bilayer structure 
displays a polar $P3$ symmetry, where Ir ions sit on a corrugated honeycomb lattice consisting of two trigonal layers intercalated by a single O layer, as shown in Figs.\ref{fig:crysband}(i) and (iii). With respect to bulk BiIrO$_3$, an additional structural distortion develops, consisting in different distances of Ir layers from the central and the interfacial O layers, the latter showing on average longer Ir-O bonds [see Fig.\ref{fig:crysband}(ii)]. The parent paraelectric (PE) structure, belonging to the $P321$ acentric (albeit non-polar, due to $D_3=2C_3+3C'_2$ symmetry) space group, is reached by tuning the offcentering of the central O layer with respect to the Ir ones, 
other distortional modes showing minor changes. Therefore, 
spin-splitting effects may appear also in the acentric PE bilayer. Due to the $C_3$ site-symmetry of Ir ions in both FE and PE structures,
a built-in bulk Rashba-like (denoted as $R$-1) and Dresselhaus-like ($D$-1) effect can be anticipated, as arising from the local dipole field and acentric environment, respectively\cite{zunger_2014}; 
in the presence of polar offcentering, additional FE-modulated $R$-1 and $D$-1 spin polarizations should appear on top of such built-in effects.

The electronic band-structure  in the valence manifold mostly arises from the upper and lower Ir-($a_{1g},\, e'_g$) orbital states, forming molecular orbitals (MOs). 
Figures \ref{fig:crysband} (a)-(d) show the two bunches of $d$ electron bands, namely 
the bonding and anti-bonding MOs arising from the $d$ orbitals at two Ir sites. 
All the 
bands are separated due to both trigonal crystal field (CF) and SOC-induced spin-splitting. The valence band maximum (VBM) is at the $K$ point; 
remarkably, all valence bands at $K$ are spin-polarized, displaying a large $s_{z}=0.41~\mu_{\rm B}$ polarization at VBM, whereas $s_{x}$ and $s_{y}$ components are  zero. 
Due to time-reversal symmetry, relating $K$ and $K'$ points,
the VBM shows a positive (negative)
spin-polarization at the $K$ ($K'$) point, a manifestation of spin-valley coupling emerging  due to the 
polar distortion, as shown in Fig.\ref{fig:spinvalley}(a). 
Interestingly, such valley-dependent spin-polarization appears to strongly depend on
the FE phase, 
being switched when the FE polarization of the heterostructure is reversed 
[Figs. \ref{fig:crysband} (b) and (d)]. 
The size of the valley-spin polarization 
evolves proportionally to the size of the Bi-O offcenter, i.e. of $P$, as shown in Fig.\ref{fig:spinvalley}(b), vanishing when $P=0$. 
In fact,  
even though a strong spin-splitting is still visible in the PE band structure shown in Fig.\ref{fig:crysband} (c), a finite $s_z$ spin polarization is found only along the $M$-$\Gamma$ line, 
being forbidden by
the $C'_2$ symmetries along the $\Gamma$-$K$ line and at the valleys $K$, $K'$.  However, a {\it hidden} valley-spin polarization can be identified when projecting the total $s_z$ onto Ir ions belonging to different layers, hinting to an intrinsic strong spin-valley coupling. 

Eventually, a complex pattern of in-plane spin polarization is shown in Fig.\ref{fig:spinvalley} (c) and (d), arising from the coexistence of several acentric distortions. 
The in-plane spin textures display the same chirality around the valleys, suggesting a substantial decoupling from the valley pseudospin. When the FE polarization is switched, a complete reversal of the spin-polarization texture is not realized, even though a clear switching of the chirality can be seen in Fig.\ref{fig:spinvalley} (c) and (d); this suggests an interplay of 
FE-dependent and built-in $R$-1 and $D$-1 effects, the latter depending on the distortional modes which are almost unchanged 
when the polar offcentering is tuned between
opposite $P_s$ and persisting
in the PE structure. 
In summary, our {\it ab-initio} calculations show a complex coexistence of valley-dependent and bulk Rashba-like (valley-independent) spin-polarization effects in bilayered BiIrO$_3$; remarkably, the spin-valley physics appears to be tightly bound to the FE distortion, emerging only in the polar structures.

The microscopic origin of the valley-spin polarization and its interplay with the FE polarization can be qualitatively grasped by considering 
a simplified tight-binding model 
 $H=H_{\rm hop}+H_{\rm cf}+H_{\rm soc}$\cite{nagaosa_2011}, describing $t_{2g}$ electrons moving on a corrugated honeycomb lattice in the presence of both SOC and trigonal CF.  We will consider here only the effects of the polar distortion --- in terms of the central O layer offcentering $\delta z$ --- that affects both Ir-O bond lengths and directions and thus determines the nearest-neighbor oxygen-mediated hopping interactions,\cite{SlaterKoster}
 neglecting the more complex structural distortions unveiled by our realistic DFT calculations. 
The Hamiltonian can be conveniently recast in a symmetry-adapted basis where the trigonal CF term becomes diagonal, while the on-site SOC 
displays the well-known expression found for $p$ ($l=1$) orbitals, the $a_{1g}$ ($e'_g$) states transforming as $p_z$ ($p_x$, $p_y$) ones\cite{SuppMat}.
Within this formulation, the analogies with graphene-like 
$p$-electron materials clearly emerge, the band-splitting induced by the trigonal CF mimicking the $sp^2$ orbitals with $\sigma, \pi$ character, whereas SOC  
may lead to Rashba-like interactions in the corrugated honeycomb lattice due to spin-mixing terms coupling the $\pi$ ($a_{1g}$) state with the $\sigma$ ($e'_g$) manifold\cite{graphene_TB1,graphene_TB2, liu_2011}. This is in striking contrast with MoS$_2$ and 
$d$-electron dichalcogenide monolayers, where spin-up and spin-down states remain completely decoupled even in the presence of SOC due to the different trigonal prismatic CF acting on $d$ orbitals (see Table \ref{tab1}). On the other hand, oxygens in the oxide bilayer provide an additional way by which inversion symmetry in the graphene-like honeycomb lattice can be broken. 
\begin{table}[th]
\caption{
Bulk symmetry (space group SG), atomic-site symmetry (site point group SPG), atomic SOC and expected spin-polarization effects  in coupled spin-valley 2D compounds. 
Following Ref. \cite{zunger_2014}, R-1 and D-1 denote Rashba and Dresselhaus effects, while R-2 and D-2 denote the  {\it hidden} spin polarizations due to the bulk IS. The electronic basis used for the  SOC term is $\bm c=(p_x, p_y, p_z)$, comprising $\sigma,\pi$ bands from $sp^2$  valence bonds, for $p$-electron systems, while it is $\bm c=(e^1_{(g)}, e^2_{(g)}, a_{1(g)})$ for $d$-electron MoS$_2$ (BiIrO$_3$ bilayer) as determined by the trigonal prismatic (octahedral+trigonal) CF.}\label{tab1}
\vspace{0.2cm}
\hspace{-0.cm}\begin{tabular}{|c|c|c|c|c|}
\hline
& {\bf SG} & {\bf SPG} & {\bf SOC} & {\bf Spin Pol.}\\
\hline
{\bf graphene} & $P6/mmm$ & $D_{3h}$ &  $\bm c_\alpha^\dagger \times \bm c_\beta^{\phantom{\dagger}}\cdot\bm \sigma_{\alpha\beta}$ & $D$-2\\
\hline
{\bf silicene} &$P\bar{3}m1$ & $C_{3v}$  & $\bm c_\alpha^\dagger \times \bm c_\beta^{\phantom{\dagger}}\cdot\bm \sigma_{\alpha\beta}$ & $R$-2 \& $D$-2\\
\hline
{\bf MoS$_2$}  & $P\bar{6}m2$& $D_{3h}$ & $(c_1^\dagger c_2^{\phantom{\dagger}}$-$c_2^\dagger c_1^{\phantom{\dagger}})\,\sigma^z$ & $D$-1\\
\hline
{\bf FE BiIrO$_3$}  & $P3$ & $C_3$  & $\bm c_\alpha^\dagger \times \bm c_\beta^{\phantom{\dagger}}\cdot\bm \sigma_{\alpha\beta}$ & $R$-1 \& $D$-1
\\
\hline
\end{tabular}
\end{table}

A standard downfolding procedure onto the $a_{1g}$ states near the $K(K')$ point allows to derive an  effective Hamiltonian for the valence bands,  $H^{\rm eff}=H_0^{\rm eff}+\delta z H_1^{\rm eff}$, 
where the polar distortion has been included up to linear order in $\delta z$ \cite{SuppMat}.  Here, $H_0^{\rm eff}$  describes the low-energy electronic properties of the ideal (non-polar) perovskite bilayer, being 
\begin{eqnarray}\label{ham:eff}
H_0^{\rm eff}&=&
t_\perp (\tau k_x S_x+k_y S_y) +\tau\sigma_z S_z\delta_0\nonumber \\&&+ \lambda_R(\sigma_x k_y-\sigma_y k_x) S_z,
\end{eqnarray}
where $\tau=\pm 1$ is a valley index labeling $K$ and $K'$, and $\bm \sigma$ and $\bm S$ denote Pauli matrices for the real spin and layer pseudospin, respectively, while $t_\perp, \delta_0$ and $\lambda_R$ are effective parameters depending on hopping amplitudes, atomic SOC and trigonal CF\cite{SuppMat}.  As anticipated before,  Eq. (\ref{ham:eff}) coincides with the low-energy model that has been derived for 2D atomic crystals with trigonal symmetry as silicene\cite{liu_2011}.
The second term describes the well-known spin-valley-layer coupling\cite{graphene_TB1,graphene_TB2, liu_2011}, 
which is responsible for the opening of a gap 
in the valence manifold.
The last term 
describes a {\it layer-dependent}  Rashba-like coupling which arises from the SOC-induced mixing of spin-up and spin-down states mediated by inter-orbital hoppings in the corrugated honeycomb lattice. 
Due to the centric symmetry of the ideal perovskite bilayer, no net spin polarization may appear; however, both the spin-valley-layer
and the layer-Rashba couplings give rise to hidden layer-dependent spin polarizations
 In the presence of the polar distortion modeled by the central O layer offcentering, 
the low-energy electronic properties are further described by 
\begin{eqnarray}\label{ham:eff1}
 H^{\rm eff}_1 &=&
E_z\,S_z+B_z \tau \sigma_z + \alpha (\tau S_x\sigma_y+S_y\sigma_x)\nonumber\\ &&+ \lambda_{R1}\left[(S_x\sigma_x-\tau S_y\sigma_y)k_y+(S_y\sigma_x+\tau S_x\sigma_y)k_x\right]\nonumber\\
&&+\lambda_R(\sigma_x k_y-\sigma_y k_x),  
\end{eqnarray} 
 where the additional effective parameters $E_z, B_z, \alpha, \lambda_{R1}$ can be expressed in terms of the tight-binding ones 
The coupling between spin and layer pseudospin described by $H_1^{\rm eff}$, removing all the degeneracies in the valence manifold, can be loosely regarded at as a valley-dependent ``magnetoelectric'' interaction\cite{gong_2013}, explaining the emergence of a net spin polarization. It is also clear that, as highlighted by the linear dependence on the polar offcentering $\delta z$, all coupling terms in Eq.(\ref{ham:eff1}) change sign upon reversal of the FE polarization, causing a global switching of the spin polarization. Specifically, the averaged {\it total} spin-polarization can be derived at valleys K (K'),  giving  $s_z=\langle\sigma_z\rangle=\tau \delta z (E_z+B_z)/\sqrt{(E_z+B_z)^2+(4\alpha)^2}$. 
 Despite the minimal parametrization of the starting tight-binding model, the derived effective Hamiltonian clearly shows that  the FE distortion is responsible for both a valley-spin polarization and an $R$-1 effect, {\it both modulated by the size of the polar distortion}.

 In conclusion, we explored the possibility of engineering the spin-valley physics in  FE transition-metal oxide heterostructures. We have explicitly shown that the interplay of trigonal CF effects and SOC in (111) oxide bilayers 
leads to graphene-like low-energy electronic properties, thus allowing for the appearance of coupled spin-valley physics. Interestingly, the spin-valley-layer coupling term can be regarded as mediating an effective valley-dependent ``magnetoelectric'' interaction which manifests in complex patterns of spin and layer polarizations;  
 when the polar distortion is induced by the oxygen-layer offcentering, 
a large valley spin-polarization, tuned by the FE polarization, develops, as we verified for a proposed candidate heterostructure comprising a robust FE BiAlO$_3$ and a bilayered BiIrO$_3$.
The appearance of valley-contrasting properties and the ``magnetoelectric'' coupling brought in by the complex interplay of valley, spin and layer pseudospin have been proposed to have important implications in valleytronic devices\cite{beenakker_2007,Xiao_2007,sakamoto_2013,gong_2013}. 
Even though the proposed material might be not for the optical and electrical generation of spin and valley polarized carriers, due to its indirect band gap and the nearly flat dispersion of the valence band maximum, we believe that our theoretical findings are more general, suggesting
that the realization of spin-valley physics in oxide heterostructures is indeed possible.
 Our theoretical analysis further suggests that such realization would allow 
in principle for larger effects (due to atomic SOC, which is typically large in $4d$ or $5d$ transition-metal ions), increased tunability (brought in by oxygens, both determining the trigonal CF splittings and mediating the hopping interactions) and for the integration of additional functionalities, such as ferroelectricity, which could be exploited in advanced next-generation electronic devices.

\acknowledgments
 This work was supported by JSPS Kakenhi (No.  22103004, 22103005 and 26800186).
P.B. and S.P. acknowledge the Italian Ministry of Research through the project MIUR-PRIN ``Interfacce di ossidi: nuove proprieta` emergenti, multifunzionalita` e dispositivi per l'elettronica e l'energia (OXIDE)''. K.Y. and P. B. contributed equally to this work.

\bibliography{biblio}

\end{document}